\documentclass[journal=jctcce,manuscript=article]{achemso}

\usepackage{amsmath}
\usepackage{amssymb}

\mciteErrorOnUnknownfalse

\usepackage{graphicx}
\usepackage{dcolumn}
\usepackage{bm}
\usepackage{lipsum}
\usepackage[T1]{fontenc}
\usepackage{mathptmx}
\usepackage{dblfloatfix}
\usepackage[usenames,dvipsnames]{color}

\title{The nature of the chemical bond in the dicarbon molecule}

\author{Claudio Genovese*}
 \affiliation{SISSA -- International School for Advanced Studies, Via Bonomea 265, 34136 Trieste, Italy;}
\author{Sandro Sorella}
\affiliation{SISSA -- International School for Advanced Studies, Via Bonomea 265, 34136 Trieste, Italy;}
\email{sorella@sissa.it}

\date{\today}

\begin{document}

\begin{abstract}
The molecular dissociation energy has often been explained and discussed in terms of singlet bonds, formed by bounded pairs of valence electrons. In this work we use a highly correlated resonating valence bond ansatz, providing a consistent paradigm for the chemical bond, where spin fluctuations are shown to play a crucial role. Spin fluctuations are known to be important in magnetic systems and correspond to the zero point motion of the spin waves emerging from a magnetic broken symmetry state. 
Within our ansatz, a satisfactory description of the carbon dimer is  
determined by the magnetic interaction of two Carbon atoms with   antiferromagnetically ordered $S=1$ magnetic moments. This is a first step that, thanks to the highly scalable and efficient quantum Monte Carlo techniques,  
may open the door  for understanding challenging complex systems containing atoms with large spins (e.g. transition metals).
\end{abstract}

\maketitle

In a non-relativistic Hamiltonian,  that do not contain explicit spin interactions, 
  the chemical interaction is, in a conventional picture,    
mostly determined  by covalent bonds where  electrons  prefer to form singlet pairs across neighbouring atoms, in this way preserving the total spin of the molecule.
In this work we will show that, rather unexpectedly,  the spin interaction 
plays a crucial role and a faithful description of the chemical bond 
can be obtained by allowing 
 local spin triplet excitations in some of the 
  electron pairs.
With  standard techniques based on molecular orbitals \cite{zensugg}, 
it is  difficult to detect this effect, and therefore we adopt the 
Resonating  Valence Bond (RVB) paradigm \cite{bk:pauling}, that has been successfully applied  
even when standard single reference molecular orbital theory fails.
Our approach  generalises the standard ''frozen singlet''
 RVB  picture, thus leading to a compact and accurate description of the electron correlation, at the prize of a very small spin contamination. 

We consider a very small molecule, but nevertheless 
very important: the carbon dimer. Its ground state is 
a perfect singlet, while the isolated atoms have two unpaired electrons, and 
any  type of mean field approach, such as HF or DFT, is completely off  
with errors of the order of eV. Also highly correlated methods,  such as 
coupled cluster \cite{doi:10.1063/1.462649,cc2008,cc2016},  face severe  difficulties in  describing 
its ground state properties, so that highly  involved  
multi-configuration expansions \cite{cc2016,doi:10.1063/1.2908237, doi:10.1063/1.3624383} are often adopted. 

In a  recent study, based on a correlated valence bond approach \cite{Shaik2012QuadrupleBI}, 
Shaik {\it et al.} have  
proposed that  a  fourth bond is necessary to explain the C$_2$ spectrum at low energy. Within  a correlated resonating  valence  bond approach, 
 they have found that the 2$\sigma$ and 3$\sigma$ molecular 
orbitals, after s-p  hybridisation, 
change  their nature as compared to standard molecular orbital theory and 
show a corresponding bonding character. By taking into account the remaining
two $\pi$ orbitals, obviously bonding, they argued that an unexpected quadruple 
bond should be a more appropriate description of the  $C_2$ molecule.
This result was rather surprising, especially considering  
that quadruple bonds should very rarely occur \cite{zhong2016latent, doi:10.1002/chem.201601382, doi:10.1002/chem.201600011, doi:10.1021/acs.jctc.6b00055, doi:10.1063/1.3555821}. 

In this work, we also adopt a RVB approach, but with the help of 
a  powerful statistical technique, named quantum Monte Carlo (QMC) \cite{mitasrmp,zenrvb}, we are able to work  with a very 
compact and accurate ansatz, that is as easy to visualise and understand as 
 a mean field ansatz. 
We consider first the most  general 
two electron pairing function:
 \begin{eqnarray} \label{pairing_gen}
 g(\vec r_1 \sigma_1, \vec r_2 \sigma_2)&=& {\frac 1  {\sqrt{2}}} (|\uparrow \downarrow \rangle  -|\downarrow \uparrow \rangle ) g_+( \vec r_1,\vec r_2)  \nonumber \\
 &+&  {\frac 1  {\sqrt{2}}} (|\uparrow \downarrow \rangle  +|\downarrow \uparrow \rangle )   g_-( \vec r_1,\vec r_2) \nonumber \\
 &+& | \uparrow \uparrow \rangle g_{\uparrow} (\vec r_1,\vec  r_2) +  | \downarrow \downarrow \rangle g_{\downarrow} (\vec r_1,\vec  r_2)
 \end{eqnarray}
where, in order to satisfy the Pauli principle $g(\vec r_1 \sigma_1, \vec r_2 \sigma_2)=-g(\vec r_2 \sigma_2, \vec r_1 \sigma_1)$, yielding 
$g_\pm (\vec r_1,\vec r_2)= \pm  g_{\pm}( \vec r_2,\vec r_1)  $
 and $g_\sigma( \vec r_2,\vec r_1)= -g_\sigma( \vec r_2,\vec r_1) $
for $\sigma= \uparrow,\downarrow$.
When considering  a generic (even) number $N$ of electrons of given spin $\sigma_i$ and positions $\vec r_i$ ($i=1,\cdots N$), 
we antisymmetrise the product 
over all the electron pairs that are determined by 
the {\em  same} pairing function. The corresponding wave function (WF) represents the most general mean-field state, namely the ground state of 
a mean-field Hamiltonian containing also BCS anomalous terms, projected onto a given number $N$ of particles and total spin projection $S^z_{tot}=\sum\limits_{i=1}\sigma_i$ along the $z$-quantization axis.
This WF  is known as the Antisymmetrised Geminal Product (AGP). 
There are three important cases: i)
when no triplet correlations are allowed, we have a perfect singlet  and we  denote it by AGPs; ii) when only  the parallel  spin 
term of  the triplet component are omitted (namely the last line in  Eq.~(\ref{pairing_gen}), the WF  can break the spin symmetry  but the magnetic order parameter can be directed only in the $z$-quantization axis, and in this case we will refer to AGPu;  iii) the most important case is the most general one 
that contains all triplet contributions. 
Henceforth  it will be indicated with AGP, as we believe it represents the most powerful description of the chemical bond within the paradigm developed in this work. Its practical implementation  requires the use of Pfaffian's algebra, as already described in Ref.~\cite{PhysRevLett.96.130201}.

\begin{figure}[t!]
        \centering
        \includegraphics[scale=0.18]{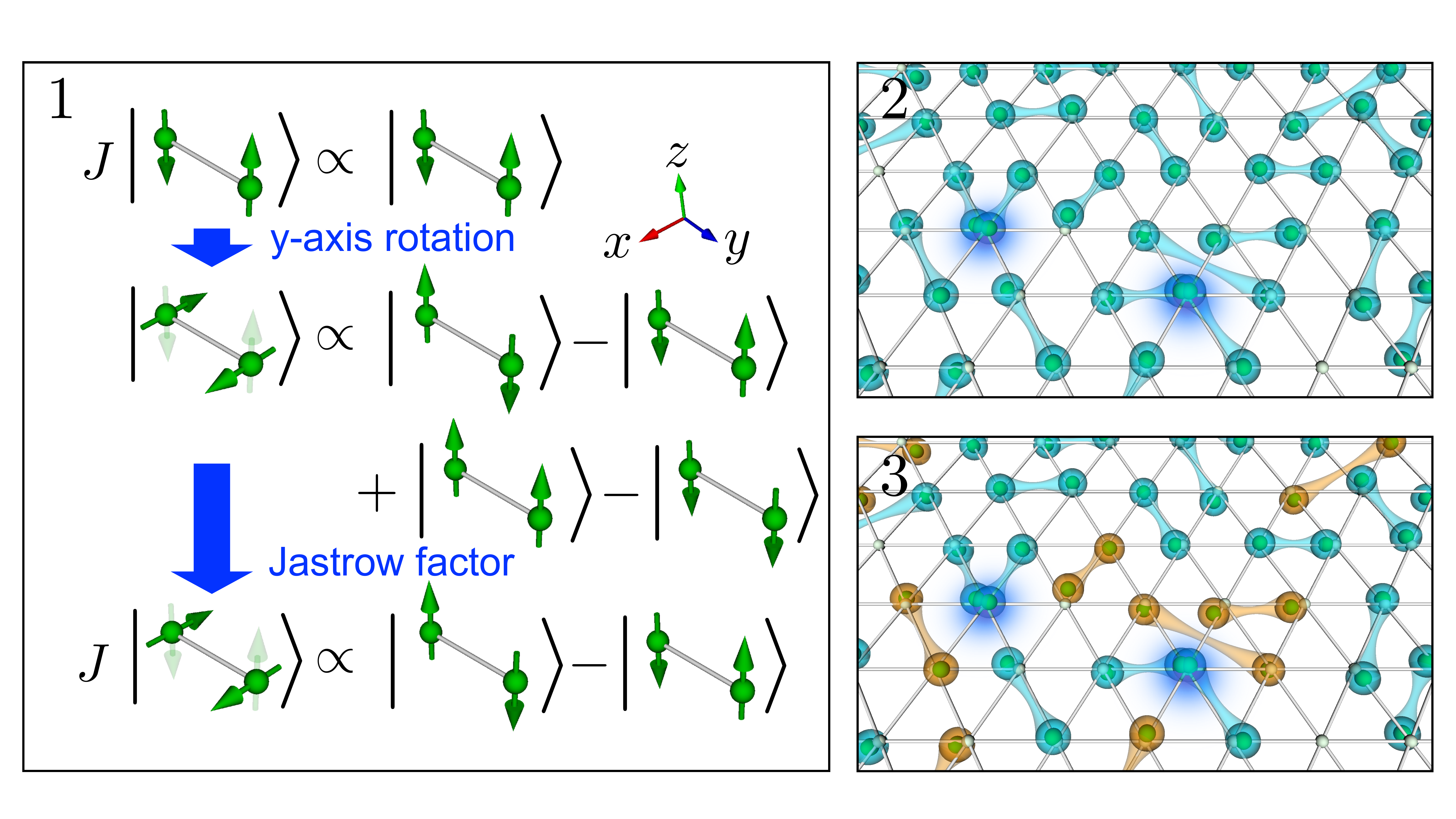}
        \caption{(1) Restoring the singlet state for the Jastrow correlated broken symmetry ansatz. 
          The Jastrow factor cannot change the broken symmetry ansatz if it is oriented in the same quantization axis ($z$-axis) of the electron basis. If we rotate the spins of the broken symmetry ansatz by 90 degrees around the $y$-axis, 
          the state becomes a quite general linear combination of spin configurations in the original basis. 
          By carefully tuning  the weights of each configuration with an appropriate 
	  spin-dependent Jastrow factor, we can recover 
	  the exact expansion of the singlet state in this basis. 
	  (2) Cartoon picture  of a typical valence bond generated  by  
	 the Jastrow correlated AGP  
	 WFs used in this work limited to only singlet bonds (JAGPs defined in main text) and 
          (3) containing both singlet and triplet bonds (JAGP).
          The balls indicate the electrons, 
          singlet and triplet bonds are displayed  with cyan and orange colours, respectively. 
          The main effect of the Jastrow factor is to reduce the probability 
	  that different electron bond pairs overlap on the same atom, in  this way avoiding the effect  of  the strong Coulomb repulsion, and dramatically improving the quality of the simpler AGP ansatz. 
        }
        \label{shirakawa1}
\end{figure}

Originally the AGPs was found to be very poor, violating size consistency even in cases where the HF was size consistent. 
The important ingredient, solving the above issues \cite{PhysRevLett.109.203001,marchimol}, 
by dramatically improving the accuracy of this WF,
is the introduction of a correlated term named Jastrow factor, that can weight 
the energetically too expensive configurations where electrons are too close and feel the large Coulomb repulsion. In this simplified picture, introduced by P.W. Anderson long time ago \cite{Anderson:1975}, one can consider with {\em a single pairing function} all the   valence bond (VB) configurations (see Figure \ref{shirakawa1}.1), in practice  with  almost optimal weights. This has been shown for instance in strongly correlated lattice models \cite{PhysRevB.62.12700}.

The Jastrow factor is an  explicitly symmetric function 
of the $N$ coordinates and spins:
\begin{equation} \label{eq:jastrow}
J(\vec r_1 \sigma_1, \vec r_2 \sigma_2 ,\cdots \vec r_N \sigma_N) = 
 \prod\limits_{i\ne j} \left[ f_{\sigma_i,\sigma_j} (\vec r_i,\vec r_j) \right],
\end{equation}
where $f$, at variance of  $g$, is the most general 
two-particle bosonic (and positive) WF. This form of the Jastrow is more general than the conventional one because, usually,  a simplified 
or even absent spin dependency in the 
$f$ function is adopted, denoted in the following with the symbol Js.
Without including spin dependency  
 a perfect singlet remains such after multiplication of the Jastrow.
However the spin dependency is necessary if we want to recover the singlet 
from a spin contaminated  broken symmetry ansatz. Though this feature is obviously 
approximate, it generalises  in a rather  convenient way the so called
``symmetry restoring 
after projection'' \cite{scuseria,PhysRevB.89.125129}.
As shown in Figure \ref{shirakawa1}.1 a simple way to understand this effect is
to  consider two atoms with opposite spins. If the spins are oriented in the 
direction of the electronic basis, defining our Jastrow in Eq.~(\ref{eq:jastrow}), the latter one cannot have any effect in the total 
WF. Conversely, it turns out energetically convenient to 
orient the spins of the atoms in a direction perpendicular to the previous one.
In this way (see Figure \ref{shirakawa1}.1)  the Jastrow can suppress the 
unfavored triplet configurations with parallel  spins.
This optimal spin-orientation   is rigorously valid within the well known 
spin-wave theory of a quantum antiferromagnet, and the 
Jastrow in this case allows the description of  the quantum 
fluctuations and the corresponding zero point energy, even for a finite 
(as is our case) number of atoms \cite{10.1143/PTP.97.399}. 

Before going in more details we summarise in Figure \ref{shirakawa1}.2-3
the physical picture of our WF in the realistic 
case when the Coulomb 
repulsion is strong and, in a given atom, it is unlikely to have 
doubly occupied orbitals. In this way the expansion of the many body WF, 
can be seen as a liquid ''soup''  of mainly  non-overlapping singlet and triplet bonds.
Our findings indicate that, despite this picture was introduced long  time 
ago,  
the important role of fluctuating triplet bonds 
has not been fully appreciated so far.

\section{Method}

After the optimization we have computed not only the total energy, 
 but also  
the total spin square \cite{doi:10.1021/ct401008s}, that can be efficiently evaluated 
by QMC.  The first quantity can be  
further improved by the so called Fixed Node projection (FN), that 
basically allows the lowest possible energy with a Jastrow $J$ not  limited only to 
two electron correlation. 
In order to avoid too cumbersome tables, in the following we report only FN 
energy  results obtained with this very accurate technique, whereas all the other quantities refer just to the bare variational  WF with the explicit  Jastrow 
written in Eq.~(\ref{eq:jastrow}).
Within this technique it is also possible 
to consider the spin square value restricted  
in a given region of space $\Lambda $, namely  
\begin{equation} \label{spin2L}  \vec S^2 ( \Lambda)= 
\left[ \int\limits_{x \in \Lambda}  d^3x \sum\limits_{i=1}^N \vec S_i \delta( \vec x - \vec r_i)  \right]^2,
\end{equation}
by extending the calculation described in Ref. \cite{genovese2020general}  only within the region $\Lambda$.
Here $\vec S_i$ is the spin operator acting on the electron at the position $\vec r_i$.
Unless otherwise specified, in all molecular diatomic calculations, $\Lambda$ is determined by  dividing the whole space into two regions separated by a plane cutting the molecule into two equivalent ones containing only one atom. 

\section{Wave Function}\label{A}
In this sectiom we will show how to define and calculate the WFs product of the general Antisymmetrised 
Geminal Power (AGP), or its restricted formulation (AGPs and AGPu), and the JF for a given configuration of electronic positions and spins
${\bm X} = \{ ({\bm r}_1 \uparrow), \cdots ({\bm r}_{N_\uparrow} \uparrow),  ({\bm r}_{N_\uparrow+1} \downarrow), \cdots ({\bm r}_N \downarrow) \}$.

\subsection{The basis set}\label{B}
The electronic coordinates are expanded in a set of localised gaussian orbitals $\left \{ \phi_{I, \nu}(\mathbf{r})  \right \}$, where $I$ and $\nu$ 
indicate the $\nu$-th orbital centred on the $I$-th atom at the  position $\mathbf{R}_I$ in the form
\begin{equation}
  \phi_{I,\nu}(\mathbf{r})=e^{-\frac{|\mathbf{r}-\mathbf{R_I}|^2} {Z_\nu}} \left[ Y_{l_\nu,m_\nu}\pm Y_{l_\nu,-m_\nu}\right],
  \label{basis}
\end{equation}
where $Z_\nu$ is a numerical coefficient that describes how diffuse the atomic orbital is around the atom, while $Y_{l_\nu,m_\mu}$ is the spherical harmonic function with angular quantum numbers $l_\nu$ and $m_\nu$. The sign of the combination $[Y_{l_\nu,m_\nu}\pm Y_{l_\nu,-m_\nu}]$ is chosen to ensure the orbital type  $\nu$ to be real. This basis set has been used without further contractions for the description of the JF, but we had to use a different solution in order to consider a large gaussian basis set for the AGP, AGPs and AGPu with a reasonable computational cost. Indeed for this study we used hybrid atomic orbitals (HO)\cite{doi:10.1063/1.1794632, doi:10.1063/1.1604379} obtained as linear combinations of elements of the gaussian basis set
\begin{equation}
  	\bar{\phi}_{I,\omega}(\mathbf{r})=\sum_\nu \mu_{\omega, \nu} \phi_{I,\nu}(\mathbf{r}).
	 \label{hyb-basis}
\end{equation}
Our choice is to use HO to describe the atomic orbitals that physically play a role. By  taking  into account  that  the atoms  considered here can be  obtained by  the occupation of  at most two s-wave and three p-wave orbitals we decided to use five HOs both for the carbon and nitrogen atoms. For the sake of compactness we indicate the basis as $\left \{\phi_{k}(\mathbf{r})  \right \}$ combining the indices $\omega$ and $I$ in a single index $k$ for a lighter notation. The use of HO basis set allows us  to expand the wave function in a large set of atomic gaussian orbitals, while remaining with a  reasonably small number of variational parameters.

\subsection{The AGP}

The AGP is the most general WF that can be built using the pairing function and contains the AGPs and the AGPu as particular cases. In the following we will explain how to calculate the full AGP including all the terms from Eq.(\ref{pairing_gen}). The algebra used for the AGP applies also to AGPs and AGPu exactly in the same way, but in the latter case a further simplification lead to the canonical determinant form for their evaluation as shown in literature \cite{doi:10.1063/1.5081933, PhysRevLett.109.203001, doi:10.1063/1.4829536, doi:10.1063/1.4829835, PhysRevB.84.245117}. 

The building block of the AGP in a given basis set $\phi(\mathbf{r})$ is, as mentioned, the pairing function whose general parametrization is
\begin{equation}
 	 g(\mathbf{r}_1  {\sigma_1},\mathbf{r}_2 {\sigma_2}) =\sum_{k,l} \lambda^{\sigma_1 \sigma_2}_{k,l} \phi_k(\mathbf{r_1}) \phi_l(\mathbf{r_2}),
 		\label{pair}
\end{equation}
where $\lambda$ is a  matrix representing  $g$ with  a finite number of variational parameters. For simplicity we will enumerate the spin up electrons from $1$ to $N_\uparrow$ and the spin down ones from $N_\uparrow+1$ to $N$. To build the AGP we consider a generic number $N$ of electrons of given spin $\sigma_i= \pm 1/2$ and positions $\vec r_i$ ($i=1,\cdots N$). We then antisymmetrize the product over all the electron pairs that,  by definition, occupy  the {\em  same} pairing function. Considering for the moment an even number of electrons we can build the matrix that contains all the pairs as
\begin{equation}
  W= \left(
  \begin{matrix}
  	0 & g(\mathbf{r}_1 \uparrow, \mathbf{r}_2 \uparrow) & \cdots & g(\mathbf{r}_1 \uparrow, \mathbf{r}_N \downarrow ) \\
  	g(\mathbf{r}_2 \uparrow, \mathbf{r}_1 \uparrow) & 0 & \cdots & g(\mathbf{r}_2 \uparrow, \mathbf{r}_N \downarrow ) \\
	\vdots & \vdots & \ddots & \vdots \\
	g(\mathbf{r}_{N-1} \downarrow, \mathbf{r}_1 \uparrow) & g(\mathbf{r}_{N-1} \downarrow, \mathbf{r}_2 \uparrow) & \cdots & g(\mathbf{r}_{N-1} \downarrow, \mathbf{r}_N \downarrow ) \\
		g(\mathbf{r}_N \downarrow, \mathbf{r}_1 \uparrow) & g(\mathbf{r}_N \downarrow, \mathbf{r}_2 \uparrow) & \cdots & 0

  \end{matrix}
  \right).
  \label{pfaff_matrix}
\end{equation}
It is a $N\times N$ matrix where to each row and to each column corresponds  an electron. We can distinguish four different blocks of this matrix depending on the spin of the pairs.
We can write the matrix $W$ as
\begin{equation}
  W= \left(
  \begin{matrix}
	W_{\uparrow \uparrow} & W_{\uparrow \downarrow} \\
        W_{\downarrow \uparrow} & W_{\downarrow \downarrow} \\
  \end{matrix}
  \right)
  \label{pfaff_matrix_small}
\end{equation}
where $W_{\uparrow \uparrow}$ and $W_{\downarrow \downarrow}$ are respectively a $N_\uparrow \times N_\uparrow $ and a $N_\downarrow \times N_\downarrow $  antisymmetric matrices that take into account the parallel spin terms of the triplet, while $W_{\uparrow \downarrow}$ is a $N_\uparrow \times N_\downarrow $ matrix such that $W_{\uparrow \downarrow} = -W^T_{\downarrow \uparrow}$, describing the remaining triplet and singlet contribution.

The extension to an odd number of electrons requires the use of an unpaired orbital that is added as last row and column of the matrix $W$. Being $\Theta_\uparrow$ the vector containing the value of $\Theta(\mathbf{r})$ calculated at the $\uparrow$ electronic positions and $\Theta_\downarrow$ the one calculated for the $\downarrow$ electronic positions.  We modify the matrix in Eq.~(\ref{pfaff_matrix_small}) as
\begin{equation}
  W= \left(
  \begin{matrix}
	W_{\uparrow \uparrow} & W_{\uparrow \downarrow} & \Theta_\uparrow \\
        W_{\downarrow \uparrow} & W_{\downarrow \downarrow} &  \Theta_\downarrow \\
        -\Theta^T_\uparrow & -\Theta^T_\downarrow & 0
  \end{matrix}
  \right).
  \label{pfaff_matrix_small2}
\end{equation}
In theory it is possible to further add an arbitrary number of pairs of unpaired orbitals  in the same way. In this case $\Theta_\uparrow$ and $\Theta_\downarrow$ are defined as matrices of size $N_\uparrow \times N_{unpaired}$ and $N_\downarrow \times N_{unpaired}$ respectively.

As suggested by the name Antisymmetrized Geminal Power, our goal is to define a WF that is literally the antisymmetrized product of the geminals and the unpaired orbitals (if present), namely 
\begin{eqnarray}
 	 \Phi(\mathbf{X}) =  \sum_\alpha  {\rm Sgn}(\alpha) \big( g(\mathbf{r}_{1_\alpha} {\sigma_{1_\alpha}},\mathbf{r}_{2_\alpha} {\sigma_{2_\alpha}})g(\mathbf{r}_{3_\alpha} {\sigma_{3_\alpha}},\mathbf{r}_{4_\alpha} {\sigma_{4_\alpha}}) \cdots  \nonumber \\
	   g(\mathbf{r}_{p-1_\alpha} {\sigma_{p-1_\alpha}},\mathbf{r}_{p_\alpha} {\sigma_{p_\alpha}})\Theta_1(\mathbf{r}_{p+1_\alpha})\cdots \Theta_{N-p}(\mathbf{r}_{N_\alpha})	 \big), \nonumber \\
	 \label{general-agp}
\end{eqnarray}
where $\alpha$ is one of the possible way of distributing the $N$ electrons between the $N/2$ pairs and the $N_{unpaired}$ unpaired orbitals $\Theta$ and ${\rm Sgn}(\alpha)$ is the sign of the corresponding permutation of the particles that is  required to insure the fermionic behaviour. The pfaffian is the algebraic operation that performs this task \cite{PhysRevLett.96.130201, PhysRevB.77.115112} yielding
\begin{equation}
  \Psi_{AGP}({\bm X})= \text{Pf}(W).
  \label{pf}
\end{equation}

\subsubsection{Pfaffian Definition}

The pfaffian is an algebraic operation defined for antisymmetric square matrix with an even number of rows and columns. This is always compatible with our matrix $W$ that is antisymmetric and has always an even number of rows and columns for each value of $N$. Before introducing the pfaffian we define a partition of the matrix $W$ as
\begin{equation}
  A(\alpha)=sign(\alpha)\prod_{k=1}^M W_{i_k,j_k}
  \label{partition}
\end{equation}
where all $i_k$ and $j_k$ are different, $i_k<j_k$ for each $k$ and $i_1<i_2<\dots<i_N$. The $sign(\alpha)$ is given by the permutation of the vector of the indices $\{i_1,j_1,i_2,j_2,\dots,i_M,j_M\}$. The pfaffian is then defined as
\begin{equation}
   \text{Pf}(W)=\sum_\alpha A(\alpha)
  \label{pf_better}
\end{equation}
where the sum over $\alpha$ is extended over all the possible partitions. The result is such that $\mbox{Pf} (W)^2=\det (W)$. To better clarify the correspondence to the Eq.~(\ref{general-agp}) an alternative definition\cite{doi:10.1063/1.1703953} of the Pfaffian can be adopted. It can indeed be defined alternatively as
\begin{equation}
	\text{Pf}(W)= \left[ (N/2)! 2^{N/2} \right]^{-1} \sum_P \text{sign}(P) \prod_{k_P=1}^{N/2} W_{i_{k_P},j_{k_P}}
	\label{kasteleyn}
\end{equation}
where $P$ now represents a generic permutation of the possible row and column indices of the matrix without any constraints and the $sign(P)$ is the parity of the permutation. It is easy to recognize the antisymmetrized sum corresponding to the Eq.~(\ref{general-agp}) in this definition. Let us introduce now a further property of the Pfaffian that will be useful for the definition of the AGPs and AGPu. In the following we will indicate with  $0$ a $m \times m$ matrix with all vanishing elements and B a generic $m\times m$ matrix,  we have that
\begin{equation}
   \text{Pf}\left[
   	 \begin{matrix}
   		0 & B\\
		-B^T & 0
   	\end{matrix}\right]=(-1)^{m(m-1)/2} \det(B).
  \label{pftoagp}
\end{equation}

\subsection {AGPs and AGPu}

In the case of AGPs and AGPu the matrices $W_{\uparrow \uparrow}$ and $W_{\downarrow \downarrow}$ are identically zero, and, in particular, for the AGPs the matrix $W_{\uparrow \downarrow}$ is symmetric. Moreover it is necessary to add $N_{unpaired}=|N_\uparrow - N_\downarrow |$  unpaired orbitals with a single spin component $\uparrow$ or $\downarrow$ depending on the polarization. Supposing for simplicity $N_\uparrow> N_\downarrow$, under these constraints, Eq.(\ref{pfaff_matrix_small2}) can be simplified as
\begin{equation}
  W= \left(
  \begin{matrix}
	0 & W_{\uparrow \downarrow} & \Theta_\uparrow \\
        -W_{\uparrow \downarrow}^T & 0 & 0 \\
        -\Theta^T_\uparrow & 0 & 0
  \end{matrix}
  \right).
  \label{pfaff_matrix_small3}
\end{equation}
With this definition we can calculate the value of the AGPs and AGPu directly using Eq.(\ref{pf}). However we can further simplify the calculation for AGPs and AGPu by noticing that Eq.(\ref{pfaff_matrix_small3}) is consistent with Eq.(\ref{pftoagp}) upon substitution of $B=(W_{\uparrow \downarrow} \Theta_\uparrow)$, yielding a determinant of a smaller matrix containing only $W_{\uparrow \downarrow}$ and $\Theta_\uparrow$.

\subsection{Jastrow Factor}

A common strategy within QMC is to multiply the WF with an exponential JF to improve the ground state description. By means of the exponential modulation provided by the JF we can take into account the electron correlation, by also  speeding up  the  convergence to the complete basis set limit. Indeed, with an appropriate choice the JF can also satisfy exactly  the electron-electron and electron-ion cusp conditions of the many-body WF, consequences  of the Coulomb $1/r$ singularity. Our JF has two main contributions
\begin{equation}
  J({\bm X})=e^{U_{ei}+U_{ee}},
  \label{JF}
\end{equation}
where $U_{ei}$ is a single body term that treats explicitly the electron-ion interaction and  $U_{ee}$ is a many-body term
to take into account the electronic correlation. The single body term is in the form 
\begin{equation}
  U_{ei}=\sum_{i=1}^{N}u_{ei}(\mathbf{r}_i),
  \label{sb1}
\end{equation}
with $u_{ei}$ being
\begin{equation}
  u_{ei}(\mathbf{r}_i)=- \sum_{I=1}^{\#ions}Z_I \frac{1-\exp(b_{ei}|\mathbf{r}_i-\mathbf{R}_I|)}{b_{ei}},
  \label{sb2}
\end{equation}
where $Z_I$ is the atomic number of the atom $I$ and $b_{ei}$ is a variational parameter.
The electron-electron term is instead written as
\begin{equation}
  U_{ee}=\sum_{i<j}u_{ee}(\mathbf{r}_i \sigma_i, \mathbf{r}_j \sigma_j),
  \label{mb1}
\end{equation}
where the sum is extended over the pairs of different electrons and where
\begin{equation}
   u_{ee}(\mathbf{r}_i \sigma_i, \mathbf{r}_j \sigma_j)=k_{\sigma_i,\sigma_j}\frac{|\mathbf{r}_i-\mathbf{r}_j|}{1+b^{ee}_{\sigma_i,\sigma_j}|\mathbf{r}_i-\mathbf{r}_j|}
+b(\mathbf{r}_i \sigma_i, \mathbf{r}_j \sigma_j),
  \label{mb2}
\end{equation}
with the $2\times 2$ matrix $b^{ee}_{\sigma,\sigma^\prime}$  described by one $b^{ee}_{\sigma,\sigma^\prime}=b^{ee}$ or two variational parameters for  $\sigma_i=\sigma_j$  when  $k_{\sigma_i,\sigma_j}=1/4$ and $b^{ee}_{\sigma,\sigma^\prime}=b^{ee}_\parallel$ and for $\sigma_i \ne \sigma_j$ when $k_{\sigma_i,\sigma_j}=1/2$ and $b^{ee}_{\sigma,\sigma^\prime}=b^{ee}_\perp$.  The conventional expression for the JF can be obtained by removing all spin dependency in  the previous expressions and remaining  only with the variational parameters corresponding to the opposite  spin case $k_{\sigma_i,\sigma_j}=1/2$ and  $b^{ee}_{\sigma,\sigma^\prime}=b^{ee}$. 

As already anticipated, thanks to the first term in Eq.~(\ref{mb2}) and the chosen  form of $u_{ei}$ we can explicitly deal with the cusp conditions of the electron-electron and electron-ion potentials, respectively. 
This highly increases the convergence to the basis set limit without considering too much sharp (with large exponents) gaussian orbitals. 
The second term in Eq.~(\ref{mb2})  instead is a bosonic pairing function $b$ that has the same definition of 
Eq.~(\ref{pair})  but without the symmetry constraints required by the fermions. The use of this term highly improves the description of the charge and 
spin correlations of the system, as opposed to the very common choice to adopt a simplified or even absent spin dependency function in the function $b$, describing only the charge correlation. In the following we have indicated the JF without spin dependency as $Js$.

\section {Results}

\begin{figure}
	\centering
	\includegraphics[scale=0.85]{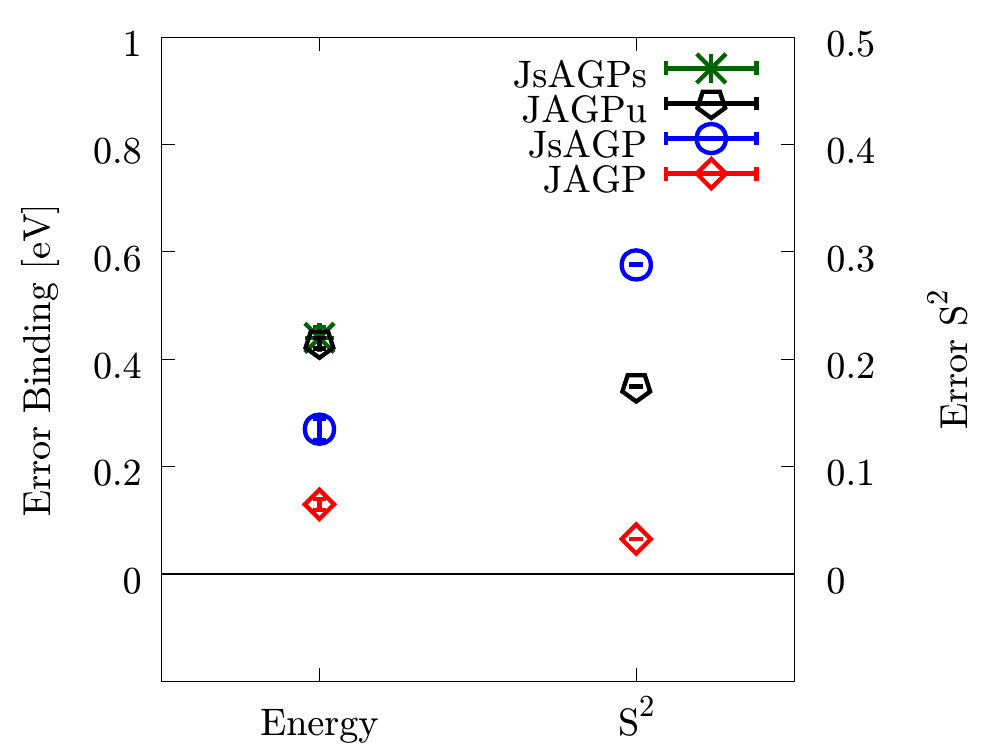}
	\caption{Energy and spin $S^2$ for different WFs.}
	\label{comparison}
\end{figure}

\begin{table}
  \caption{Spin measures with the  different WFs.}\label{tab:spinsquare}
  \begin{tabular}{l c c c c}
    \hline
    \hline
    \rule[-0.4mm]{0mm}{0.4cm}
    &  \multicolumn{3}{c}{$S^2$} & $2 \mu_B$ \\
    \rule[-0.4mm]{0mm}{0.4cm}
    &  Atom & Molecule & Half Molecule & Moment $\parallel z$\\
    \hline
    \rule[-0.4mm]{0mm}{0.4cm}
    JsAGPs & 2.00 & 0.00 & 1.529(4) & 0.0005(4)\\
    \rule[-0.4mm]{0mm}{0.4cm}
    JAGPu  & 2.00534(3) &  0.1743(5)  & 1.760(2) & 0.5833(4)  \\
    \rule[-0.4mm]{0mm}{0.4cm}
    JsAGP  & 2.00418(5) &  0.2880(4)  & 2.0185(5)  & 0.7194(4)\\
    \rule[-0.4mm]{0mm}{0.4cm}
    JAGP  & 2.00542(1) &  0.0327(1)  & 2.0649(3) & 0.0013(5) \\
    \rule[-0.4mm]{0mm}{0.4cm}
    Exact  & 2.00 & 0.00 & -  & - \\
    \hline
    \hline
\end{tabular}			
\end{table}

In the following we report detailed calculations by using the setup and the methodology described in the previous sections. As it is shown in Tab.~\ref{tab:spinsquare}, in the broken symmetry ansatz JsAGPu
the  total spin value is rather 
large, implying a large spin contamination, 
that is not present in the atom.
By turning on the triplet correlations in the pairing function 
$g$ and restricting the Jastrow factor to depend only on the electronic 
densities, i.e. the JsAGP case in Tab.~\ref{tab:spinsquare}, the contamination  is even worse with an increase of the total spin. 
Remarkably,  by  using the proposed  spin-dependent 
Jastrow factor J in Eq.~(\ref{eq:jastrow}), 
it is possible to recover an almost  exact singlet WF.
This confirms the picture that the C$_2$ molecule can be considered as 
the smallest  antiferromagnet made of two atoms with opposite spins. Indeed, when we go from the broken symmetry ansatz 
and include correctly the quantum fluctuations in $J AGP$ the WF becomes almost exact, because, as it is shown in Figure \ref{comparison}
, it is not only an almost perfect singlet but  recovers almost all the molecular binding. 
The  energy associated with spin fluctuations can  be quantified in 
analogy with  the simple model of two $S=1$ spins  
interacting with an Heisenberg 
antiferromagnetic coupling  $J \vec S_1 \cdot \vec S_2$\cite{book} .
Here the singlet ground state with energy $-{2 J }$ is below 
the classical energy $-J$ of two antiparallel spin along  a chosen direction. 
Analogously  the JAGP ansatz has  a spin fluctuation energy gain 
referenced to the corresponding JAGPu one (see Fig.~\ref{shirakawa1}.1).
As it is shown in the in Figure (\ref{comparison}) and in appendix Tab.II 
this energy gain 
 represents a large fraction of  the estimated energy ($\simeq 0.5eV$\cite{Shaik2012QuadrupleBI})  to break the fourth bond, i.e. $\simeq 0.30 eV$.

\begin{figure}[t]
	\centering
	\includegraphics[scale=0.25]{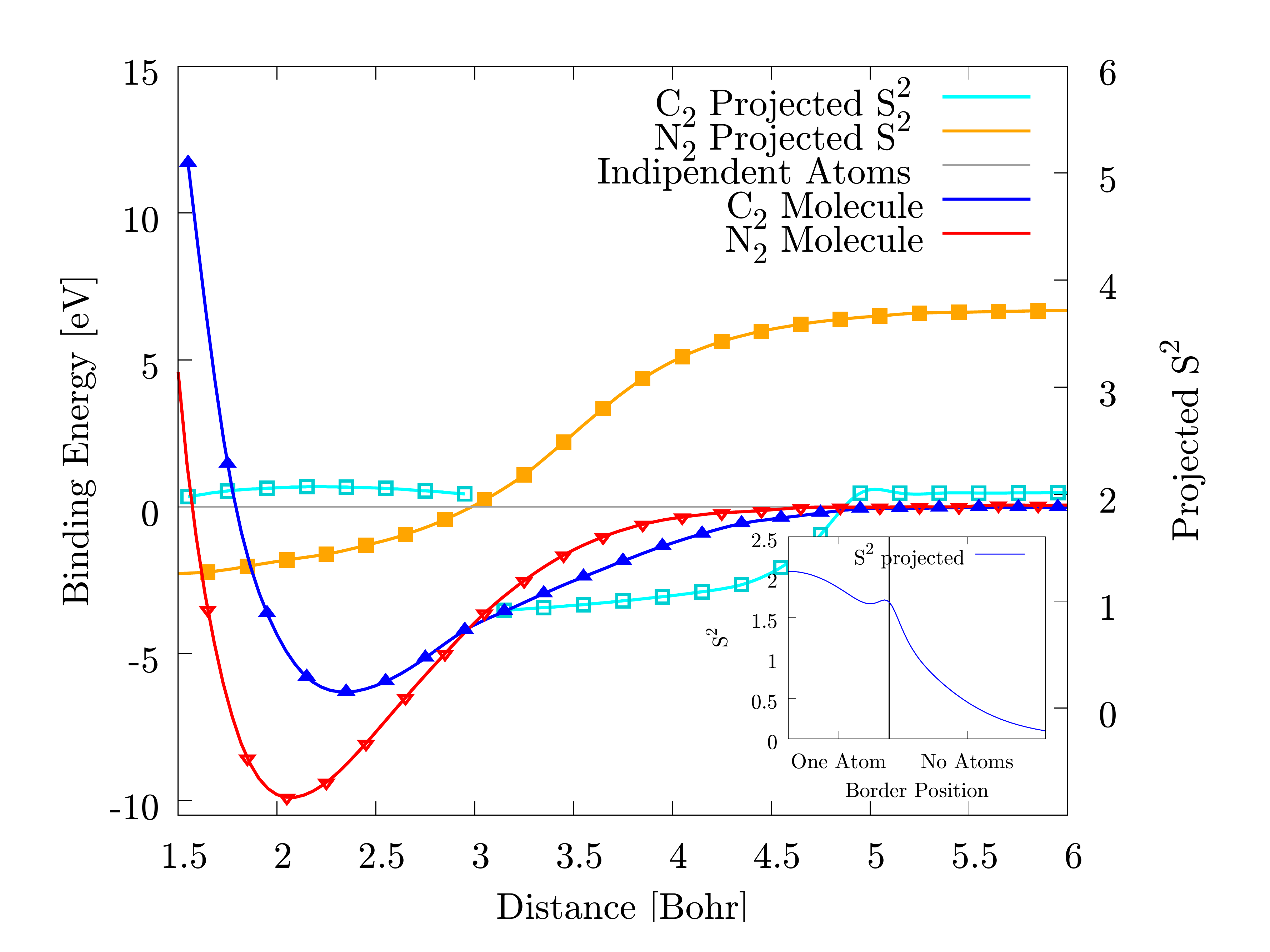}
	\caption{Size consistency of carbon and nitrogen dimers: both the systems at large distance recover the energy 
and the expectation value of the $S^2$ operator
        of two isolated atoms. At bond distance however the carbon atoms maintain a large localized $S^2$, while the spin of the nitrogen atoms is much smaller than the corresponding atomic value. Lines are guides to the eye.  
	In the carbon dimer the sharp change  of the projected $S^2$ at  around $3\ a.u.$ distance is probably a consequence of an avoided crossing between two energy levels with the same $^1\Sigma^+_g$ symmetry, that has been shown  in an accurate and detailed  DMRG  calculation\cite{doi:10.1063/1.4905237}. In the inset the projected spin square [Eq.~(\ref{spin2L}) in the text] of the C$_2$  molecule at equilibrium  computed as a function of $\Lambda$. The semi-infinite region  contains  one or no atom as  a function of the distance between the plane perpendicular to the molecular axis, defining this region, and the center of the molecule.}
	\label{sizeconsistency}
\end{figure}

\begin{table}
  \caption{Expectation value of the total spin square $S^2$ for different systems. 
In this table also an evaluation  of the local  moment   $S_{half}^2$ 
is shown, that is defined here by 
 the expectation value of the spin  square  operator, measured in the two semi-infinite regions, each one containing half the molecule, specularly  symmetric with  respect to  a plane perpendicular to the molecular axis.}\label{tab:spincomp}
  \begin{tabular}{l c c}
    \hline
    \hline
    \rule[-0.4mm]{0mm}{0.4cm}
    &  {$S^2$} & {$S_{half}^2$} \\
    \hline
    \rule[-0.4mm]{0mm}{0.4cm}
    $C_2$   & 0.0327(1)   & 2.0649(3)  \\
    \rule[-0.4mm]{0mm}{0.4cm}
    $N_2$   & 0.0061(1)  & 1.3760(3)  \\
    \rule[-0.4mm]{0mm}{0.4cm}
    $HCCH$   & 0.0045(1)  & 1.3889(2)  \\
    \hline
    \hline
\end{tabular}			
\end{table}

When  the two atoms   are placed at large distance it is clear that their spins have to be  equal to the corresponding isolated $S=1$ atomic value.
What is really surprising in this small molecule, 
is that (see Figure \ref{sizeconsistency}) the spin $S^2(\Lambda)$ around a single atom is almost the same ($S^2(\Lambda)=2$) when we are at bond distance, showing that the local spin moment is large even in this regime. 
This effect is very robust (see  the inset in Fig.~\ref{sizeconsistency})  as we have  
found that almost all the atomic spin is extremely localized around the nuclear positions, and its value can be measured with no ambiguities.
This is a special feature of the C$_2$ molecule, and contrasts with the behaviour of other molecules. 
For instance there exists  an ongoing debate on why 
the chemical bond in $C_2$ is stronger than in acetylene, which has a triple bond, but $C_2$ has a longer bond and a smaller force constant than the C-C bond of HCCH. In our picture the simple answer is that the C2 chemical bond 
is increased by  a 
sizable ''long distance'' 
magnetic contribution because the atomic spin moment is 
basically unscreened as opposed to a standard triple bond molecule, triple bond that is obviously more effective to reduce the bond length, dominated by charge electrostatic.
 Indeed in Tab.~\ref{tab:spincomp} it is evident 
 that in the triple bond $N_2$ and 
 $HCCH$ dimers  the total spin square $S^2_{half}$ calculated for each monomer  is much smaller than the one expected for an independent $N$ atom or a $CH$ group, suggesting that the high value of $S^2_{half}$ for the carbon dimer is a peculiar feature of this system. Moreover in Fig.~\ref{sizeconsistency} we report the JAGP dispersion curve for the N$_2$ molecule: in a standard 
triple bond, the spin around each atom is substantially screened and therefore 
much below the free atomic value, fully recovered only at large distance.
Note that also the N$_2$  molecule  is very well  represented by our ansatz, e.g. with an estimated well depth of $229.1(1)$ kcal/mol in very good agreement with  the estimated exact one of $228.48(6)$ kcal/mol \cite{doi:10.1063/1.1869493}.

As shown  in the appendix,  our variational estimate is essentially the state of  the art for an all-electron calculation.

\begin{figure}[t]
	\centering
	\includegraphics[scale=0.24]{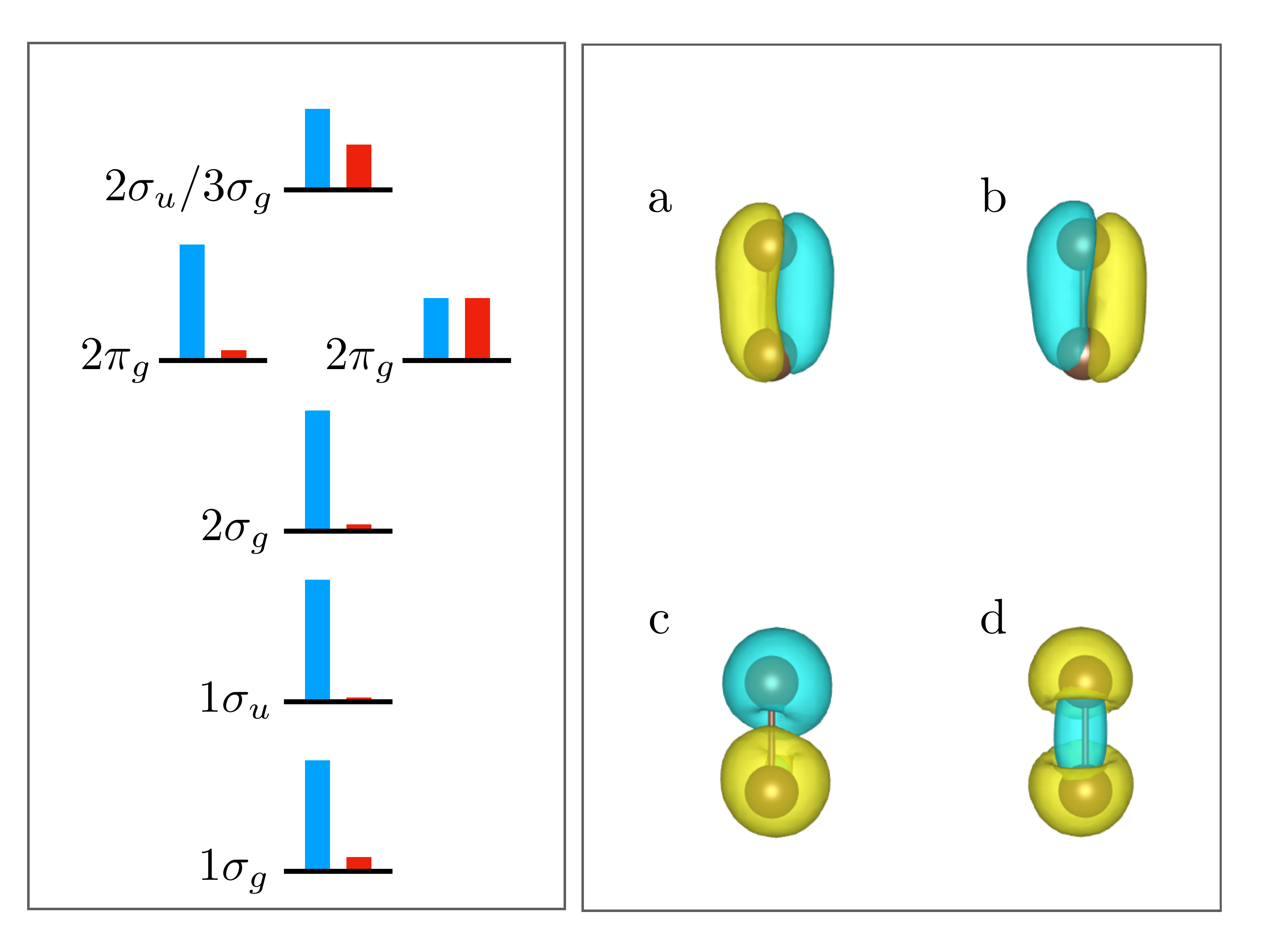}
	\caption{Left panel: after breaking the spin symmetry, each spin-independent orbital of an unrestricted Slater determinant ansatz splits into a pair of single occupied ones with no definite spin projection.
 The histograms represent the corresponding  spin component weights: the height of the blue (red) rectangle indicates the percentage of the majority (minority)  spin.  Notice that the occupation order is different in DFT calculations where the order is $1\sigma_g/1\sigma_u/2\sigma_g/2\sigma_u/2\times(2\pi_g)$ and where the orbitals have a single spin component. Right panel:  majority and minority spin orbitals with the most relevant spin contamination. The orbitals (a) and (b) come from the $2\pi_g$ and have the same weight, the orbitals (c) and (d) are the last occupied ones (indicated as $2\sigma_u/3\sigma_g$), in this case the (c) orbital has a $65\%$ and the (d) one a $35\%$ weight.}
	\label{orbitals}
\end{figure}

\section{Conclusion}
We have studied in details the simple C$_2$ molecule and found, rather unexpectedly over previous expectations, that a large fraction of the chemical bond can be attributed to spin fluctuations. Indeed, when we allow our WF for triplet 
correlations both in the Jastrow and in the mean field part, 
we obtain a remarkable improvement in the description of the bond.

It is particularly instructive  to see the role of correlation  in  modifying  
the molecular orbitals, by taking as a reference the ones  corresponding  to a simple DFT double bond picture. The AGP part of our WF, after full optimization of 
the energy in presence of the Jastrow factor, can be also recasted in terms of 
molecular orbitals by  an almost  standard diagonalization\cite{genovese2020general}.
The spin character of the resulting 
orbitals are displayed in Fig.\ref{orbitals}, where it is shown that i) at variance of the DFT mean-field ansatz,  the $3\sigma_g$ bonding 
orbital is eventually 
present as the minority spin component of  the 
HOMO, in partial agreement with the quadruple bond picture.  ii)
in our ansatz however the main effect that mostly determines the chemical
bond is the spin contamination of the most important occupied orbitals 
$2\sigma_u$,$3\sigma_g$ and $2\pi_g's$.
In  this way they can contribute to the bonding by means of the corresponding spin-fluctuation energy gain, that is instead vanishing for the inner core orbitals (this is because they have a  definite spin in  the same  quantization  axis chosen for the Jastrow factor,  see explanation in Fig.~\ref{shirakawa1}.1).

We have shown therefore  that 
the bonding in $C_2$  cannot be explained with charge electrostatic, and instead the large atomic spin value confirms that the energy is intimately due to correlation, the same that allows, by means of our Jastrow factor, the evaluation of the spin-wave zero point energy of a quantum antiferromagnet.


\section *{Acknowledgements}
We thank A. Zen and K.  Jordan for useful  discussion and for carefully reading the manuscript  before submission.
We acknowledge PRIN-2017  for financial support  and CINECA PRACE-2019 for computational resources.

\section *{Data Availability}
The data that support the findings of this study are available from the corresponding author upon reasonable request.

\bibliography{Bibliography}

\appendix
 
 \section{Total Energy and Excitations}
In table \ref{tab:TOTenergies} we compare the FN energies of our JAGP WF with the ones obtained with other state of the art variational methods.

\begin{table}[h!]
    \caption{Total energy comparison between different variational methods and WFs.}\label{tab:TOTenergies}
    \begin{tabular}{l l}
      \hline
      \hline
      \rule[-0.4mm]{0mm}{0.4cm}
      WF & Total Energy [H]\\  
      \hline
      \rule[-0.4mm]{0mm}{0.4cm}
      JsAGPs & -75.8938(2) \\  
      \rule[-0.4mm]{0mm}{0.4cm}
      JAGPu &  -75.8935(2))\\  
      \rule[-0.4mm]{0mm}{0.4cm}
      JSD${}^a$ &  -75.8672(1) \\  
      \rule[-0.4mm]{0mm}{0.4cm}
      JAGP & -75.9045(2)\\  
      \rule[-0.4mm]{0mm}{0.4cm}
      JFVCAS${}^a$  & -75.9106(1) \\
      \rule[-0.4mm]{0mm}{0.4cm}
      FCIQMC${}^b$ & -75.80251(8) \\
      \hline
      \hline
      \multicolumn{2}{l}{${}^a$ Reference \cite{doi:10.1063/1.2908237}.}\\
      \multicolumn{2}{l}{${}^b$ Reference \cite{doi:10.1063/1.3624383}.}\\
    \end{tabular}			
  \end{table} 

Moreover, we also report in Tab.~\ref{spectrum}, the direct and adiabatic triplet excitations of the molecule, obtained within the JAGP. We recover very well the small energy gap to the $^3 \Pi$ excitation, that is highly challenging. Another important excitation is the vertical  $^3 \Sigma^+$ state, that was used by Shaik {\emph et al.} to estimate the binding energy of the quadruple bond. 
This is also correctly estimated by our ansatz and represents the lowest energy ''spin-wave'' in this simple antiferromagnetic system.

  \begin{table}
    \caption{Carbon dimer excitations.}\label{spectrum}
    \begin{tabular}{l c c c }
      \hline
      \hline
      \rule[-0.4mm]{0mm}{0.4cm}
      & \multicolumn{3}{c}{$X^1\Sigma^+$} \\
      \rule[-0.4mm]{0mm}{0.4cm}
      & Energy [H] & Binding [eV] & Gap [eV]\\  
      \hline
      \rule[-0.4mm]{0mm}{0.4cm}
      JAGP & -75.9045(2)  & 6.31(1)  & -  \\
      \rule[-0.4mm]{0mm}{0.4cm}
      Estimated Exact${}^a$ & -75.9265 & 6.44  & - \\
      \rule[-0.4mm]{0mm}{0.4cm}
      Experiment${}^b$ & -  & 6.30  & - \\
      \hline
      & & & \\
      \rule[-0.4mm]{0mm}{0.4cm}
      &   \multicolumn{3}{c}{$a^3\Pi$} \\
      \rule[-0.4mm]{0mm}{0.4cm}
      &  Energy [H] & Binding [eV] & Gap [eV]\\  
      \hline
      \rule[-0.4mm]{0mm}{0.4cm}
      JAGP & -75.8961(2) & 6.08(1)  & 0.23(1)  \\
      \rule[-0.4mm]{0mm}{0.4cm}
      Experiment${}^c$ & -  & - &  0.09 \\
      \hline
      & & & \\
      \rule[-0.4mm]{0mm}{0.4cm}
      & \multicolumn{3}{c}{$b^3\Sigma^+$} \\
      \rule[-0.4mm]{0mm}{0.4cm}
      &  Energy [H] & Binding [eV] & Gap [eV]\\  
      \hline
      \rule[-0.4mm]{0mm}{0.4cm}
      JAGP & -75.8680(3) & 5.30(2) & 0.99(2)  \\
      \rule[-0.4mm]{0mm}{0.4cm}
      Experiment${}^c$ & -  & -  & 0.80 \\
      \hline
      & & & \\
      \rule[-0.4mm]{0mm}{0.4cm}
      & \multicolumn{3}{c}{$b^3\Sigma^+$ Vertical} \\
      \rule[-0.4mm]{0mm}{0.4cm}
      &  Energy [H] & Binding [eV] & Gap [eV]\\  
      \hline
      \rule[-0.4mm]{0mm}{0.4cm}
      JAGP &  -75.8658(1) & 5.26(1) & 1.05(1)  \\
      \rule[-0.4mm]{0mm}{0.4cm}
      Experiment${}^d$ & - & -  & 1.14 \\
      \hline
      & & & \\
      \rule[-0.4mm]{0mm}{0.4cm}
      &  \multicolumn{3}{c}{$c^3\Sigma^-$}\\
      \rule[-0.4mm]{0mm}{0.4cm}
      &  Energy [H] & Binding [eV] & Gap [eV]  \\  
      \hline
      \rule[-0.4mm]{0mm}{0.4cm}
      JAGP & -75.8737(2) & 5.47(1)  & 0.84(1)  \\
      \rule[-0.4mm]{0mm}{0.4cm}
      Experiment${}^c$ & - & -  & 1.04 \\
      \hline
      \hline
        \multicolumn{4}{l}{${}^a$ Reference \cite{doi:10.1063/1.1869493}.}\\
        \multicolumn{4}{l}{${}^b$ Reference \cite{nist}.}\\
        \multicolumn{4}{l}{${}^c$ Reference \cite{doi:10.1021/cr00098a005}.}\\
        \multicolumn{4}{l}{${}^d$ Reference \cite{article1234521}.}\\
    \end{tabular}			
  \end{table}

\end{document}